\documentclass[prb,aps,showpacs,floatfix,floats,nobalancelastpage,twocolumn]{revtex4-1}
\usepackage{graphicx}
\usepackage{graphics}
\usepackage{latexsym,amsmath,amssymb,bm,euscript}
\usepackage{color}
\usepackage{dcolumn}
\usepackage{bm}
\usepackage{epstopdf}
\usepackage{times}
\usepackage{multirow}
\usepackage{wasysym}
\usepackage{subfigure}

\def\ra{\rangle}
\def\la{\langle}

\def\Hc{{\rm H.c.}}

\def\rhombpic{{\begin{picture}(26,15)(-2,-2)
                     \put (12,0) {\circle*{5}}
		     \put (24,0) {\circle*{5}}
		     \put (6,10) {\circle*{5}}
		     \put (18,10) {\circle*{5}}
		     \put (12,0) {\line (1,0) {12}}
		     \put (6,10) {\line (1,0) {12}}
                     \put (18,10){\line(3,-5){6}}
		     \put (6,10) {\line (3,-5) {6}}
               \end{picture}}}
               
\def\linepic{{\begin{picture}(12, 0)(0,0)
                    \put(0,0) {\circle*{5}}
                    \put(12,0) {\circle*{5}}
                    \put(0,0){\line (1,0) {12}}
                  \end{picture}}}

\def\triangpic{{\begin{picture}(17,15)(-2,-2)
                      \put (0,0) {\circle*{5}}
		      \put (12,0) {\circle*{5}}
		      \put (6,10) {\circle*{5}}
		      \put (0,0) {\line (1,0) {12}}
		      \put (12,0) {\line (-3,5) {6}}
		      \put (0,0) {\line (3,5) {6}}
                \end{picture}}}

\begin{document}

\title{ Possible spin liquid states with parton Fermi surfaces in the SU(3) ring-exchange model on the triangular lattice}
\author{Hsin-Hua Lai}
\affiliation{National High Magnetic Field Laboratory, Florida State University, Tallahassee, Florida 32310, USA}
\date{\today}

\begin{abstract}
We consider a SU(3) ring-exchange model on a triangular lattice. Unlike the SU(2) case, under perturbation expansion of the SU(3) Hubbard model, the three-site ring exchange is present as well as the usual four-site ring exchange. Interestingly, the three-site ring exchange differs from the usual two-site and four-site exchanges by a minus sign and is ferromagnetic. We first present numerical site-factorized state studies on this model which shows a three-sublatticed order phase and a ferromagnetic phase. We further study the model using slave-fermion mean field in which we rewrite the exchange operators in terms of three flavors of fermions. We find the main competing trial states are the trimer state (triangular plaquette state) and the gapless U(1) spin liquid states with parton Fermi surfaces which include both the uniform zero-flux spin liquid state, and the uniform $\pi$-flux spin liquid state. Furthermore, we find there are possible pairing instabilities of the zero-flux (Fermi surface) spin liquid state toward a $f$-wave gapless (nodal) spin liquid state and the $\pi$-flux (Fermi surface) spin liquid state toward an interesting exotic $s$-wave ``gapless'' spin liquid state with two flavors of fermions paired up while one flavor of fermions remains gapless.
\end{abstract}
\maketitle

\section{Introduction}\label{Introduction}
In recent years, the cold atomic systems have become a powerful tool to realize strongly-correlated systems, including N-flavor Hubbard model on different lattices. Such systems consisting of several flavors of interacting fermions can be realized as different hyperfine states of alkali atoms \cite{Honerkamp2004} or nuclear spin states of ytterbium \cite{Cazalilla09, Fukuhara2009, Taie2010} or alkaline-earth atoms \cite{CWu03, Gorshkov10, Tey2010}. A model Hamiltonian to describe such systems is the N-flavor fermionic Hubbard model \cite{Honerkamp2004, Xu2010, Gorshkov10}
\begin{eqnarray}
H = - t \sum_{\la jk \ra}\sum_{\alpha} \left[ c^{\alpha\dagger}_j c^{\alpha}_k + \Hc \right] + U \sum_{j}\sum_{\alpha,\beta}n^{\alpha}_j n^{\beta}_j,~
\end{eqnarray}
where $\alpha,~\beta$ run over the different flavors, $\la jk \ra$ runs over pairs of nearest neighbors on the lattice, and $j$ runs over all lattice sites.

The most common case is when $N=2$, which corresponds to the usual spin-$1/2$ Hubbard model. If we focus on the half-filling condition, i.e., when each site is occupied by exactly one fermion, the system undergoes metal-to-Mott insulator phase transition for sufficiently large repulsion $U$. In experiments on cold atoms, the Mott insulating state of fermions have recently been observed at the low temperature compared with the Fermi temperature. \cite{Robert08, Schneider2008} In the large $U$ regime, it is generally accepted that the ground state is well-captured by the usual anti-ferromagnetic (AFM) Heisenberg model and the charge degrees of freedom are completely frozen out. However, in the regime where $U$ is not sufficiently large compared with the hopping strength $t$, the ground state is not well-understood. It is possible that the strong charge fluctuations play important role for stabilizing ``gapless'' spin liquid phases in this regime.\cite{ringxch,LeeandLee05,HYYang2010}

For the more general case with $N > 2$, \cite{Affleck1988, Read1989, Read1990, Harada2003, Naoki2007, Beach2009, Hermele2009, Hermele2011, Cai2012} if we focus on certain fillings, the Mott insulating states will also emerge. In this case, the spin order is not understood even in the large $U$ limit in which we can only focus on the Heisenberg-like (two-site exchange) Hamiltonian 
\begin{eqnarray}
H_{2} = J\sum_{\la jk \ra} P_{jk} ,~
\end{eqnarray}
where $P_{jk}$ is so-called two-site exchange operator, which permutes the fermions between two nearest-neighbor sites as $P_{jk} | \alpha,~ \beta \ra = | \beta,~ \alpha \ra$, where the $\alpha,~\beta$ represent the spin states at sites $j$ and $k$. For $N =3$, there has been numerical evidence on such SU(3) Heisenberg model on a triangular lattice suggesting three-sublattice ordered ground state in this regime. \cite{Bauer2012}

The situation becomes more complex if $t/U\sim O(1)$, since in this regime the two-site exchange Hamiltonian is not sufficient to capture the essential physics and higher ordered contributions such as ring exchanges should be taken into account. In this paper, we focus on this regime and consider the SU(3) ring-exchange model on the triangular lattice. Motivated by the perturbative studies of the SU(3) Hubbard model at $1/3$ filling, we include the ``FM`` three-site ring exchanges and the ``AFM`` four-site ring exchanges in the model whose Hamiltonian is given in Eq.~(\ref{SU(3)_H}). With the interplay between the ring exchanges, we conjecture that the Quantum spin liquid (QSL) states \cite{LeeNagaosaWen,Balents_nature} can arise due to the strong frustration.

In this work, we first study the ordered phases using the site-factorized ansatz.\cite{Lauchli2006} We find that the phase diagram contains the three-sublattice ordered phase, similar to the phase found in Ref.~\onlinecite{Bauer2012}, and the FM state. We further study this model using slave-fermion trial states. After performing numerical full optimization of the trial energy, we consider the three ansatz states such as the uniform zero-flux and $\pi$-flux gapless (Fermi-surface) spin liquid state and the trimer (plaquette) state. Furthermore, we consider several pairing instabilities and find that there is a pairing instability of the zero-flux spin liquid state toward a $f$-wave (nodal) spin liquid state; there is a pairing instability of the uniform $\pi$-flux spin liquid state toward a interesting exotic $s$-wave spin liquid state with two flavors of fermions paired up while one flavor of fermions remains gapless. 

The paper is organized as follows. In Sec.~\ref{Sec:model} we define explicitly the model Hamiltonian we will study in the paper. In Sec.~\ref{Subsec:site_factor} we use the site-factorized ansatz to study the ordered states in this ring-exchange model. In Sec.~\ref{Subsec:MFcalc} we use the slave-fermion representation to rephrase the SU(3) Hamiltonian in terms of the three-flavor fermionic Hamiltonian and perform the fermionic mean-field treatment of the model. We further study the possible pairing instability in the spin liquids regime. In Sec.~\ref{Sec:Discussion} we conclude with some discussions.

\section{SU(3) Model with ring exchange terms}\label{Sec:model}
\begin{figure}[t]
\includegraphics[width=\columnwidth]{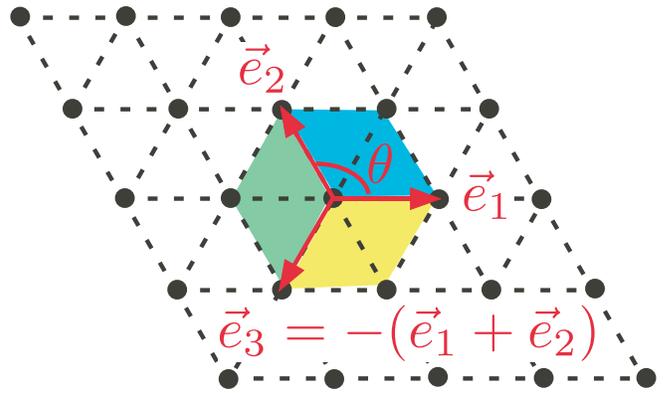}
\caption{The triangular lattice showing the vectors $\vec{e}_{\nu=1,2,3}$ used in the text. The three different colored regions represent the three rhombi associated with the site at the vector center and we label each site counterclockwise from 1 to 4. The angle $\theta$ between each vector $\vec{e}$ is $2 \pi/3$.  
}
\label{triangular_lattice}
\end{figure}

The model Hamiltonian we consider is 
\begin{eqnarray}\label{SU(3)_H}
\nonumber H_{SU(3)} = && J \sum_{\linepic} P_{12} - K_{3} \sum_{\triangpic}  \left[ P_{123} +\Hc \right] \\
&& + K_4 \sum_{\rhombpic} \left[ P_{1234} + \Hc \right],~~~~
\end{eqnarray}
with $~\linepic~$ running over all the bonds on the lattice;$~\triangpic~$ running over all the triangles, up- and down-triangles, on the lattice, while $\la 123 \ra$ are the sites on the triangles labeled counterclockwise;$\rhombpic~~$ running over all the rhombi ( for one site, there are three rhombi associated with it--blue, green and yellow shaded rhombi, see Fig.~\ref{triangular_lattice}) and $\la 1234 \ra$ are the sites on a rhombus labeled counterclockwise. $P_{12}$ is the nearest-neighbor two-site exchange operator, $P_{123}$ is the three-site spin ring exchange operator, which permutes the fermions on the triangles as $P_{jkl} |\alpha, \beta, \gamma \ra = |\gamma, \alpha, \beta \ra$, and $P_{1234}$ is the four-site spin ring exchange operator, with $P_{jklm} |\alpha, \beta, \gamma, \eta \ra = | \eta, \alpha, \beta, \gamma \ra$, Fig.~\ref{triangular_lattice}. The couplings $J$, $K_3$, and $K_4$ can be obtained from perturbative analysis of SU(3) Hubbard model at $1/3$ filling and the leading-order terms are 
\begin{eqnarray}
&& J=\frac{2t^2}{U}, \\
&& K_3 = \frac{6t^3}{U^2}, \\
&& K_4 = \frac{20 t^4}{U^3}.
\end{eqnarray}
Without the ring exchange terms, previous studies of the site-factorized ansatz on the triangular lattice predicted a three-sublattice ordered state \cite{Papanicolaou1988, Tsunetsugu2006, Lauchli2006} which was recently confirmed by Density Matrix Renormalization Group (DMRG) and infinite Projected Entangled-Pair States (iPEPS) analysis \cite{Bauer2012}. 

Recently, Ref.~\onlinecite{Bieri2012} did variational studies on the SU(3) model with three-site ring exchange, and they found the FM state and the three-sublattice ordered state in the regime of FM three-site ring exchanges. On the AFM side of the three-site ring exchanges, they found an interesting $d_x + i d_y$ spin liquid state with a gapless parton Fermi surface. 

However, the situation is not clear if both the three-site and four-site ring exchange terms are included. In this section, we will focus on the regime where the three-site ring exchange is FM and the four-site ring exchange is AFM, motivated by the perturbative studies of SU(3) Hubbard model. In order to study the ordered phases, below in Sec.~\ref{Subsec:site_factor}, we first present our studies using the site-factorized states. Later in Sec.~\ref{Subsec:MFcalc} we will present our studies using the slave-fermion trial states.

\subsection{Site-factorized state studies}\label{Subsec:site_factor}

\begin{figure}[t]
\includegraphics[width=\columnwidth]{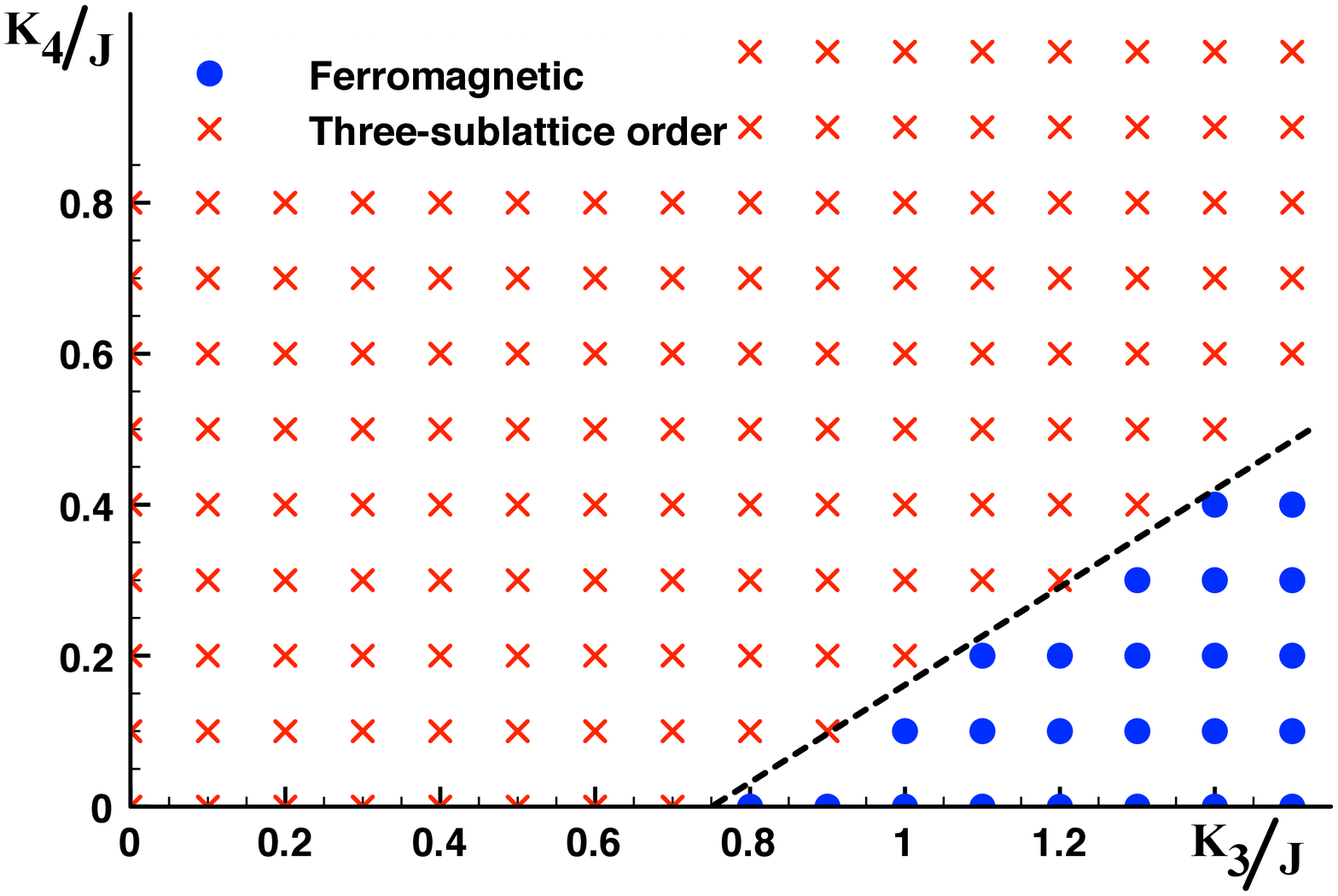}
\caption{
The phase diagram using the site-factorized states. The red crosses represent the three-sublattice ordered states with each on-site vector, Eq.~(\ref{site-factorized-vect}), mutually orthogonal to each other and the closed blue circles represent the FM state. The dashed line is the exact boundary between the two phases, see texts. The boundary between the FM state and the three-sublattice ordered state can be determined analytically (see text) and consistent with the numerical studies. We note that this model contains a four-sublattice ordered state for strong four-site ring exchange, roughly when $K_4/J >1.4$. In such state, the numerical site-factorized states indeed show a four-sublattice periodicity. However, the four-sublattice ordered state is hard to be characterized analytically.
}
\label{site_factorized_pdg}
\end{figure}

In this subsection, we consider the site-factorized state \cite{Lauchli2006} defined as
\begin{eqnarray}\label{site-factorized-vect}
|s \ra = \prod_{j} | \mathcal{X}_j \ra,
\end{eqnarray}
with
\begin{eqnarray}
| \mathcal{X}_j \ra \equiv a_{j} | x \ra_{j} + b_{j}  | y\ra_{j} + c_j  | z \ra_{j},
\end{eqnarray}
where we fix the overall phase by setting the phase of $a_j$ to be zero such that $a_j \in~\mathbb{R}$ and $b_j,~c_j~\in~\mathbb{C}$ and $|a_j|^2 + |b_j|^2 + |c_j|^2 =1$. Above, we used the usual time-reversal invariant basis of the SU(3) fundamental representation \cite{Lauchli2006}, defined as
\begin{equation}\label{xyz_transf}
\begin{array}{ccc}
|x \ra = \frac{i | 1 \ra - i | -1 \ra}{\sqrt{2}}; & |y\ra =\frac{|1 \ra + | -1 \ra}{\sqrt{2}}; & |z \ra =-i | 0 \ra,
\end{array}
\end{equation}
with $|S^z = \pm 1 \ra \equiv |\pm1\ra$ and $|S^z = 0\ra \equiv |0\ra$. According to the parametrization of the $|\mathcal{X}_j\ra$ vector along with the constraint, at each site there are $4$ independent parameters. For a lattice with $N\times N$ sites, there are $4N^2$ independent parameters for the site-factorized state. We numerically calculate the optimized (lowest) site-factorized state energy, $E_{sf} = \la s| H_{SU(3)} |s \ra$, on a $3\times3$, and on a $6\times6$ triangular lattice with periodic boundary condition using the gradient descent method. In the site-factorized state studies, we find the energies of the optimal states do not change upon the system size.

The phase diagram is shown in Fig.~\ref{site_factorized_pdg}. The red crosses represent the states with three-sublattice order with sublattices labeled as $A,~B,$ and $C$. The three-sublatticed states are characterized with each on-site vector ($|\mathcal{X}_{j\in A}\ra, |\mathcal{X}_{j \in B}\ra,$ and $|\mathcal{X}_{j \in C}\ra$) mutually orthogonal to each other and the closed blue circles represent the FM state. We only show the phase diagram of magnetic ordered states up to parameter regime of $O(K_3/J)\sim O(K_4/J)\sim 1$, because the quantum fluctuations would become more important upon increasing the strength of the ring exchanges, and the magnetic ordered phases are more likely to be destroyed. The conjecture is confirmed by the linear flavor wave theory calculation discussed in the discussion, Sec.~\ref{Sec:Discussion}. We note that this model shows four-sublattice ordered state with strong four-site ring exchange, roughly when $K_4/J > 1.4$. In the regime, the numerical values of the site-factorized states show a four-sublattice periodicity. However, the four-sublattice ordered state is hard to be characterized analytically.\cite{Mark:helical_state, Toth2010} In addition, we also numerically check $4\times 4$ triangular lattice and find the four-sublattice ordered state, but its energy is higher than the three-sublattice ordered state in the present parameter regime, which is consistent with our studies on $6\times 6$ system.

The boundary between the three-sublattice order and the ferromangetic phase can be understood analytically. The three-sublattice state is a state with each on-site vector mutually orthogonal to each other. The state vector for the three-sublattice state has the form $|\psi \ra_{u} = |x\ra_{j\in A} \otimes |y\ra_{k\in B} \otimes |z\ra_{l \in C}$ with $j,~k,~l$ around a triangle. Since the state vector at each site is orthogonal to each other, the energy is $\la H_{SU(3)} \ra_{3-sub} = 0$. On the other hand, the state vectors of a FM phase are aligned with each other, say $|z\ra$ for all sites. The energy in the FM phase is $\la H_{SU(3)} \ra_{FM} = 3 J - 4 K_3 + 6 K_4$. The boundary between these two phases can be determined analytically from the condition $3J - 4 K_3 + 6 K_4 = 0$. For $K_4 = 0$, the transition point is at $3/4$, which is consistent with our numerical studies. 
\subsection{Slave-fermion trial states and energetics}\label{Subsec:MFcalc}
In this subsection, we follow the approach similar to the one outlined in Ref.~\onlinecite{Serbyn2011} for the spin $S = 1$. We write the spin operators in terms of three flavors of fermionic spinons, $f^{\alpha}$,
\begin{equation}\label{Rep:flavor}
\begin{array}{lr}
S^{\alpha}_j = - i \sum_{\beta, \gamma} \epsilon^{\alpha \beta \gamma} f^{\beta \dagger}_j f^{\gamma}_{j},&~~~~~ \sum_{\alpha}f^{\alpha \dagger}_{j} f^{\alpha}_{j} = 1,
\end{array}
\end{equation}
with $\alpha,~\beta,~\gamma \in \{x, y, z\}$ and $j$ is the site label. We note that the fermionic spinon $f$ is different to the electron $c$ in the Hubbard model. Conceptually, we are focusing on the insulating side of the metal-to-Mott insulator phase transition, in which the charge degrees of freedom are localized. In the mean field level, one assumes that the spinons do not interact with one another and are hopping freely on the 2D lattice. The mean field Hamiltonian would have the spinons hopping in zero magnetic field, and the ground state would correspond to filling up a spinon Fermi sea. In doing this one has artificially enlarged the Hilbert space, since the spinon hopping Hamiltonian allows for unoccupied and doubly-occupied sites, which have no meaning in terms of the spin model of interest. It is thus necessary to project back down into the physical Hilbert space for the spin model, restricting the spinons to single occupancy.

The fermion representation $\{f^x, f^y, f^z\}$ in Eq.~(\ref{Rep:flavor}) can be related to the usual fermion representation $\{f_{+1}, f_{-1}, f_{0}\}$, where $ f_{\pm 1}$ carry $S^z$ quantum number $\pm 1$ and $f_{0}$ carries $S^z$ quantum number $0$, based on Eq.~(\ref{xyz_transf}):
\begin{equation}
\begin{array}{ccc}
f^x = \frac{-i}{\sqrt{2}}\left[f_{+1} - f_{-1}\right],& f^y = \frac{1}{\sqrt{2}}\left[f_{+1} + f_{-1}\right],
& f^z = i f_0.
\end{array}
\end{equation}
In $\{f_{\pm1}, f_{0}\}$ basis, the spin operator can be represented as
\begin{eqnarray}
&& S^{+} \equiv S^x + i S^y = \sqrt{2} \left( f^\dagger_{+1} f_0 + f^\dagger_0 f_{-1}\right),\\
&& S^z = f^\dagger_{+1} f_{+1} - f^\dagger_{-1} f_{-1}.
\end{eqnarray}

The exchange operators in terms of fermions are
\begin{eqnarray}
&&P_{jk} = \sum_{\alpha \beta} f^{\alpha \dagger}_{j} f^{\beta}_{j} f^{\beta \dagger}_{k} f^{\alpha}_{k},\\
&& P_{jkl} = \sum_{\alpha \beta \gamma} f^{\alpha \dagger}_{j} f^{\beta}_{j} f^{\beta \dagger}_{k} f^{\gamma}_{k} f^{\gamma \dagger}_{l} f^{\alpha}_{l}, \\
&& P_{jklm} =\sum_{\alpha \beta \gamma \eta} f^{\alpha \dagger}_{j} f^{\beta}_{j} f^{\beta \dagger}_{k} f^{\gamma}_{k} f^{\gamma \dagger}_{l} f^{\eta}_{l} f^{\eta \dagger}_{m} f^{\alpha}_{m},~ 
\end{eqnarray}
where $\alpha, \beta, \gamma, \eta = x,~y,~z$.

The Hamiltonian, Eq.~(\ref{SU(3)_H}), can be re-expressed as
\begin{eqnarray}
\nonumber H_{SU(3)} =&& J \sum_{\linepic} \sum_{\alpha \beta}  f^{\alpha \dagger}_{1} f^{\beta}_{1} f^{\beta \dagger}_{2} f^{\alpha}_{2} - \\
\nonumber  && \hspace{-0.5cm} - K_3 \sum_{\triangpic} \sum_{\alpha \beta \gamma} \bigg{[} f^{\alpha\dagger}_1 f^{\beta}_{1} f^{\beta \dagger}_{2} f^{\gamma}_2 f^{\gamma \dagger}_3 f^{\alpha}_{3} + \Hc \bigg{]} + \\
&&\hspace{-1cm} + K_4 \sum_{\rhombpic}\sum_{\alpha \beta \gamma \eta}  \bigg{[}  f^{\alpha \dagger}_1 f^{\beta}_1 f^{\beta \dagger}_2 f^{\gamma}_2 f^{\gamma \dagger}_3 f^{\eta}_3 f^{\eta \dagger}_4 f^{\alpha}_4  + \Hc \bigg{]}.~~~~~~
\end{eqnarray}
Below, we will calculate the trial energies using the slave fermion trial states. We start from the case without considering pairing instability. We find that the main competing states are what we call the ``trimer'' state, the uniform $\pi$-flux spin liquid state, and the zero-flux spin liquid state. Later, focusing on the regime in which spin liquid states are the optimal slave-fermion states, we consider the possible pairing instabilities. We find that there is a possible pairing instability of the zero-flux spin liquid state toward a $f$-wave (nodal) spin liquid state and a pairing instability of the $\pi$-flux spin liquid state toward an exotic $s$-wave spin liquid state with two flavors of fermions paired up while one flavor of fermions remains gapless.
\subsubsection{Without pairing instability}\label{Subsubsec:nopairing}
In this section, we focus on the non-magnetic trial states. When we perform numerical calculations, we relax the constraint of the fermion number for each flavor to be
\begin{eqnarray}\label{constraint}
\la f^{\alpha \dagger}_j f^{\alpha}_j \ra_{trial}= \frac{1}{3}.
\end{eqnarray} 
A convenient formulation of the mean field is to consider a general SU(3)-rotation invariant trial Hamiltonian
\begin{eqnarray}\label{MF_H:nopairing}
\nonumber H_{trial} = && - \sum_{\la jk \ra}\sum_{\alpha} \left[ t_{jk}e^{- i \theta_{jk}} f^{\alpha \dagger}_{j} f^{\alpha}_{k} + \Hc \right] -\\
&& -\sum_{j}\sum_{\alpha} \mu_{j}f^{\alpha \dagger}_{j} f^{\alpha}_j,~
\end{eqnarray}
with  $t_{jk}$ being the hopping amplitude, $\theta_{jk}$ being the phase of the hopping $t_{jk}$ in different mean-field ansatz states, and $\mu_j$ being the chemical potential which can be used to satisfy the constraint, Eq.~(\ref{constraint}). With the trial Hamiltonian above, we can find the ground state and use it as a trial wave function for the Hamiltonian $H_{SU(3)}$, Eq.~(\ref{SU(3)_H}). After performing ``complete'' Wick contractions and ignoring the constant pure density terms, the trial energy can be expressed as
\begin{widetext}
\begin{eqnarray}\label{MF_energy:nopairing}
\nonumber E_{MF} = && -J \sum_{\linepic} \bigg{|} \sum_{\alpha} \chi^{\alpha}_{12} \bigg{|}^2 - \\
\nonumber && - K_3 \sum_{\triangpic} \bigg{\{}\bigg{[} \sum_{\alpha} \chi^{\alpha}_{12}  \chi^{\alpha}_{23} \chi^{\alpha}_{31}  - \sum_{\alpha \beta} \bigg{(} n^{\alpha}_{1} \chi^{\alpha}_{23} \chi^{\beta}_{32} +  n^{\alpha}_{2} \chi^{\alpha}_{31} \chi^{\beta}_{13} + n^{\alpha}_{3} \chi^{\alpha}_{12} \chi^{\beta}_{21} \bigg{)} +  \sum_{\alpha \beta \gamma} \chi^{\alpha}_{13} \chi^{\beta}_{32} \chi^{\gamma}_{21} \bigg{]} + \Hc \bigg{\}} + \\
\nonumber && + K_4 \sum_{\rhombpic} \bigg{\{} \bigg{[} \sum_{\alpha} \bigg{(} n^\alpha_2 \chi^{\alpha}_{13} \chi^{\alpha}_{34} \chi^{\alpha}_{41} + n^\alpha_4 \chi^{\alpha}_{12} \chi^{\alpha}_{23}  \chi^{\alpha}_{31} \bigg{)} - \\
\nonumber && \hspace{1.5cm}  - \sum_{\alpha \beta} \bigg{(} n^\alpha_1  n^\alpha_2  \chi^\alpha_{34} \chi^{\beta}_{43} + n^\alpha_1  n^\alpha_4  \chi^{\alpha}_{23} \chi^{\beta}_{32} + n^{\alpha}_3  n^{\alpha}_4  \chi^{\alpha}_{12} \chi^{\beta}_{21} +n^{\alpha}_{2}  n^{\alpha}_{3}  \chi^{\alpha}_{41} \chi^{\beta}_{14} + n^{\alpha}_2  \chi^{\alpha}_{31} n^{\beta}_{4}  \chi^{\beta}_{13} + \chi^{\alpha}_{12} \chi^{\alpha}_{34} \chi^{\beta}_{23}  \chi^{\beta}_{41} \bigg{)} + \\ 
\nonumber && \hspace{1.5cm}+ \sum_{\alpha \beta \gamma} \bigg{(} \chi^{\alpha}_{12} \chi^{\alpha}_{34} \chi^{\beta}_{21}  \chi^{\gamma}_{43} + \chi^{\alpha}_{23} \chi^{\alpha}_{41} \chi^{\beta}_{32} \chi^{\gamma}_{14} + n^{\alpha}_{2}  \chi^{\alpha}_{31} \chi^{\beta}_{14}  \chi^{\gamma}_{43} + n^{\alpha}_{4}  \chi^{\alpha}_{13} \chi^{\beta}_{21} \chi^{\gamma}_{32} \bigg{)} - \\
 && \hspace{1.5cm} - \sum_{\alpha \beta \gamma \eta} \chi^{\alpha}_{14} \chi^{\beta}_{43} \chi^{\gamma}_{32} \chi^{\eta}_{21} \bigg{]} + \Hc \bigg{\}}. 
\end{eqnarray}
\end{widetext}
Above we defined $( \chi_{jk}^\alpha)^* \equiv \la f^{\alpha \dagger}_j f^{\alpha}_k \ra_{trial}$. 

The slave-fermion trial states which conserve the translational symmetry we consider in this work are the uniform zero-flux state and uniform $\pi$-flux state, which represents the spin liquid states with uniform $\phi=\theta_{jk}=0$ and $\phi=\theta_{jk}=\pi$ respectively for all nearest $<jk>$. The fermions in both of these trial states hop isotropically on the lattice, $t_{jk} \simeq t$, and therefore the expectation values of $\chi_{jk}$ are expected to be isotropic, $\chi^{\alpha}_{jk} \simeq \chi^{\alpha} = const$. Below we list the numerical values of the energy per site in the two states 
\begin{eqnarray}
&& E^{MF}_{\phi=0} = -0.7672 J + 0.4482 K_3 -0.4521 K_4, \label{MF_zeroflux:num_val}\\
&& E^{MF}_{\phi=\pi} = -0.4395 J +0.8352 K_3 - 0.7949 K_4. \label{MF_piflux:num_val}
\end{eqnarray}
We can observe that the optimal translationally invariant slave-fermion state favors $\phi=0$ state when $K_3 >0$ and $\phi=\pi$ state when $K_3 <0$. The distinguishability between these two uniform-flux spin liquid states can be related to the origin of the $K_3$ which arises from the third-order perturbation of the SU(3)Hubbard model at $1/3$ filling. However, $K_4$ does not distinguish these two uniform-flux (U(1)) spin liquid states from this perspective. 

\begin{figure}[t]
\includegraphics[width=\columnwidth]{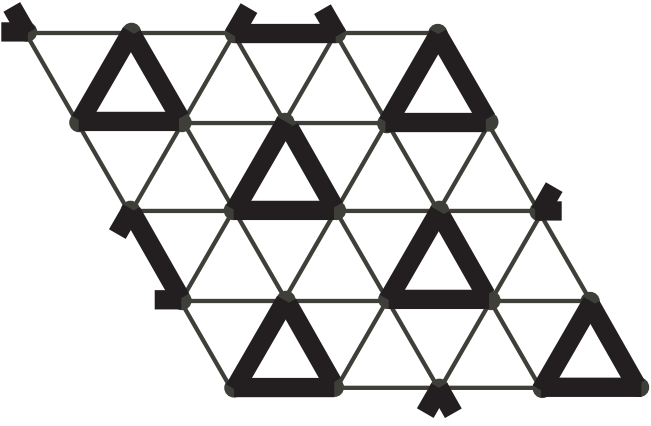}
\caption{Illustration of the trimer state. In slave fermion picture, the fermionic spinons only hop around each triangular plaquette and we can focus on each triangle separately.}
\label{trimer}
\end{figure}

Besides the translationally invariant state, we also consider what we call the ``trimer'' state. Fig.~\ref{trimer} shows one example of the configuration of such a state in which the non-zero $t_{jk}$ form non-overlapping trimer covering of the lattice. These states break translational invariance, and any trimer covering produces such a state. Such states can have lower Heisenberg exchange energy. The occupied bonds attain the maximal expectation value which is found analytically $|\chi^{\alpha}_{jk}|_{max} = n^{\alpha}_j = 1/3$. Their contribution can be sufficient to produce the lowest total energy and such states are expected to be the lowest-energy states with $K_3 = 0$ and $K_4 = 0$. 
\begin{eqnarray}
E^{MF}_{trimer} = - J + 0.5926K_3 - 0.3704 K_4.~\label{MF_trimer:num_val}
\end{eqnarray}
The energies of different slave-fermion trial states are functions of $K_3/J$ and $K_4/J$ and it is expected that different optimal trial state is realized in different parameter regime. In order to clearly show the cross between the energies of different mean-field ansatz states as functions of $K_3/J$ and $K_4/J$, we plot Eqs.~(\ref{MF_zeroflux:num_val})-(\ref{MF_trimer:num_val}) with $J \equiv 1$ in the limit of either $K_3=0$ or $K_4 =0$. 

\begin{figure}[t]
\subfigure[]{\label{Plots4diff_MF_K4=0} \includegraphics[width=\columnwidth]{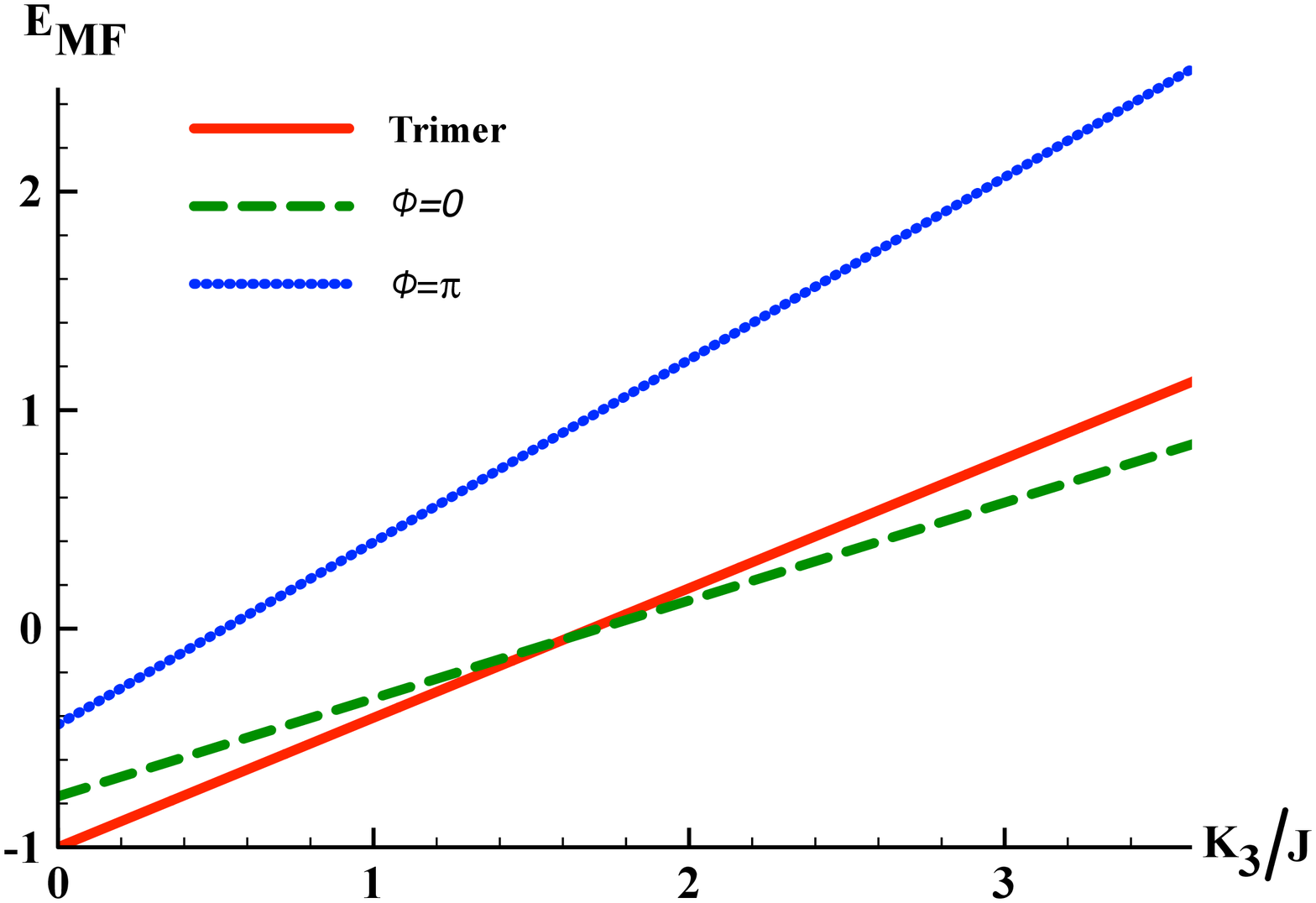}}
\subfigure[]{\label{Plots4diff_MF_K3=0}\includegraphics[width=\columnwidth]{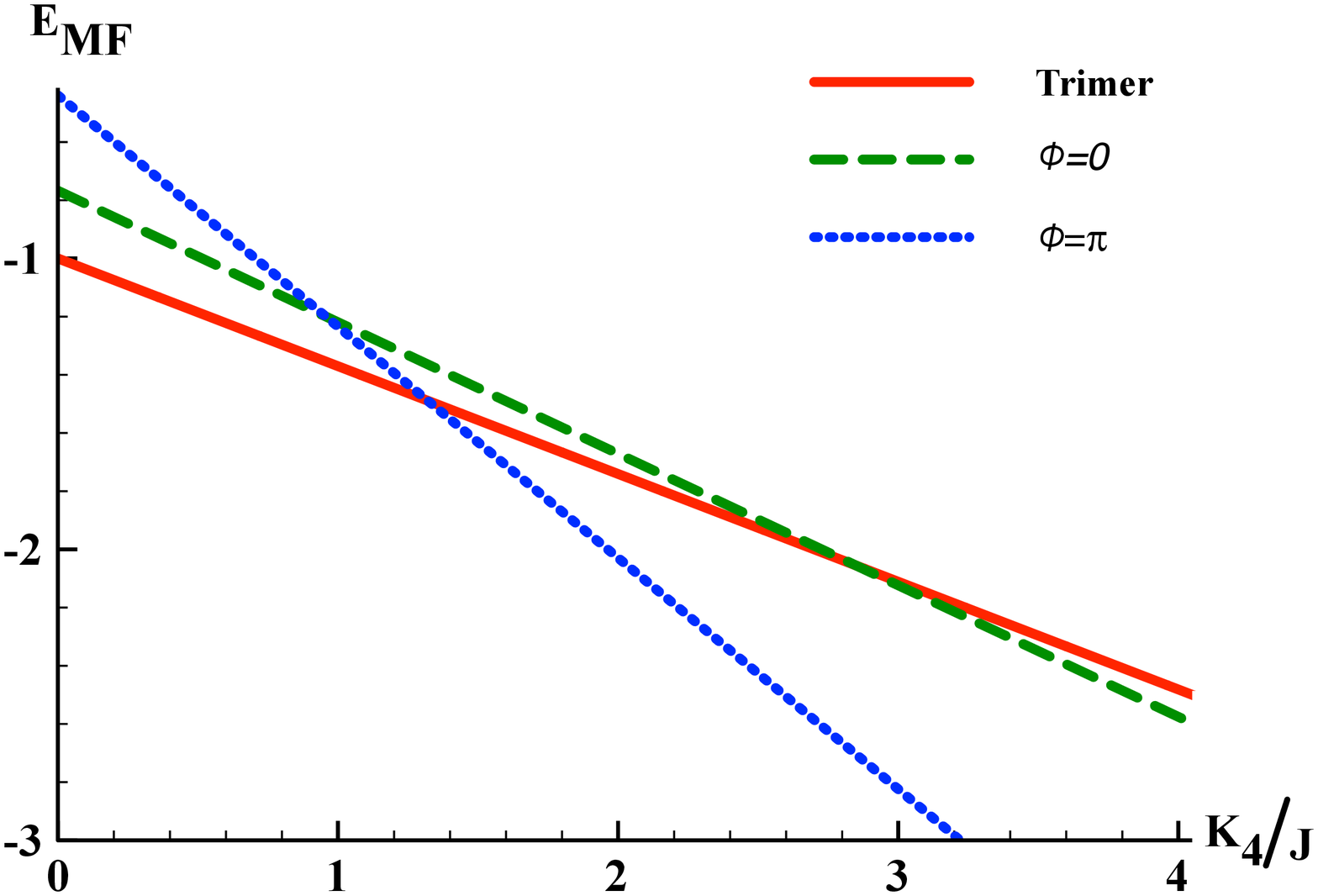}}
\caption{(a) Illustration of energies of different slave-fermion trial states as a function of $K_3$ with $K_4 =0$ and $J\equiv 1$ (b) Illustration of energies of different slave-fermion trial states as a function of $K_4$ with $K_3=0$. When $K_3 = 0$ and $K_4 =0$, the optimal state is the trimer state followed by the zero-flux spin liquid state and the $\pi$-flux spin liquid state. When $K_3$ increases while $K_4 =0$, the zero-flux spin liquid state becomes the lowest-energy state at $(K_3  >1.61, K_4=0)$ as shown in (a). On the other hand, when $K_4$ increases while $K_3 =0$, the energy line of $\pi$-flux spin liquid state first crosses the energy line of the zero-flux spin liquid at $(K_3 =0, K_4\simeq0.96)$ and then crosses the energy line of trimer state to become the lowest energy state at $(K_3 =0, K_4\simeq 1.32)$. For general cases, the complete mean-field phase diagram is shown in Fig.~\ref{MFSL_pdg} }  
\label{Plots4diff_MF_E}
\end{figure}

\begin{figure}[t]
\includegraphics[width=\columnwidth]{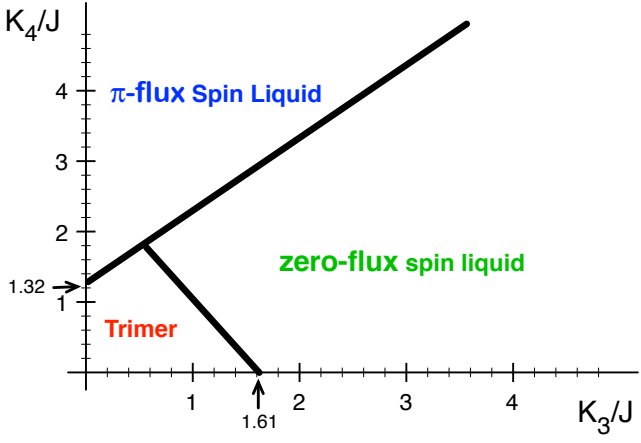}
\caption{ The phase diagram of the mean-field ansatzes. The lines are the tentative boundaries between different phases. We note that the zero-flux and $\pi$-flux spin liquid states are both U(1) Fermi-surface spin liquid states. The difference is that the zero-flux spin liquid state only possesses a single Fermi pocket in the center while the $\pi$-flux spin liquid state possesses two Fermi pockets near the hexagonal Brillouin zone. The exact wave function of the trimer state can be written down explicitly, Eq.~(\ref{trimer:wf}), and we can calculate the corresponding energy exactly.
}
\label{MFSL_pdg}
\end{figure}

Figure~\ref{Plots4diff_MF_E} shows the energies of different mean-field states as a function of $K_3/J$ with $K_4 =0$  and as a function of $K_4/J$ with $K_3 =0$. For the former, Fig.~\ref{Plots4diff_MF_K4=0} clearly shows that at $K_3,~K_4 = 0$ the trimer state is the lowest energy state followed by the uniform zero-flux state and $\pi$-flux state. When $K_3$ is gradually increased, the energy line of the zero-flux state crosses that of the trimer state and the zero-flux state becomes the lowest energy state at $K_3/J \sim 1.61$. On the other hand, for the latter, Fig.~\ref{Plots4diff_MF_K3=0} shows that the energy line of $\pi$-flux state first crosses that  of the zero-flux state at $K_4/J\simeq 0.96$ and then crosses that of trimer state. The zero-flux state becomes the lowest energy state at $K_4/J \simeq 1.32$. For a complete phase diagram, we numerically determine the ground states in the $K_3-K_4$ parameter regimes and the result is summarized in the mean-field phase diagram, Fig.~\ref{MFSL_pdg}.

In order to check if the mean-field ansatz states we considered are sufficient to describe the physics in this model, we perform numerically ``full optimization'' of the mean-field energy, Eq.~(\ref{MF_energy:nopairing}), on a triangular lattice with $100 \times 100$ 3-site unit cells, by treating $\chi_{jk}$-s and $\theta_{jk}$-s as varying variables. In the numerical optimization, there are in total $18$ variables, $9$ $\chi_{jk}$ and $9$ $\theta_{jk}$, and we take $t_{jk} =1$, $\mu^{\alpha}_j = \mu$. Numerics suggest that the above trial states are the three optimal states.

Before leaving this section, we want to remark that the trimer state is a singlet state around a triangular plaquette, and we can write down the exact singlet wave function  in a closed form as 
\begin{eqnarray}\label{trimer:wf}
|\psi_{trimer}\ra = \sum_{\alpha, \beta, \gamma}\frac{\epsilon^{\alpha \beta \gamma}}{\sqrt{6}} | \alpha, \beta, \gamma \ra, 
\end{eqnarray}
with $\alpha=x,y,z$. With the trimer wave function, we can calculate the energy per site 
\begin{eqnarray}
\mathcal{E}_{\psi_{trimer}} = -\frac{1}{3} J -\frac{4}{27}K_3 + \frac{2}{9} K_4.
\end{eqnarray} 
For $K_3 =0$ and $K_4=0$, the exact trimer state energy is very close to the variational energy of U(1) spin liquid which is $-0.34 J$\cite{Bieri2012}, but much higher than the three-sublattice order state energy obtained by DMRG which is roughly $-0.678 J$. \cite{Bauer2012} However, such a plaquette state can be stabilized with finite $K_3$ before reaching the FM state. To see this, we can compare the energy of the FM state as shown analytically in the end of Sec.~\ref{Subsec:site_factor} with that of the exact trimer state energy. Knowing the energy of the FM state, $E_{FM} = 3J - 4K_3+6K_4$, we can see that the trimer state indeed has the lower energy than that of the FM state when $K_3 < 45/52+3K_4/2\simeq 0.86+1.5 K_4$ with $J\equiv1$. 
\subsubsection{With pairing instability}\label{Subsubsec:pairing}
So far, we have ignored the possible pairing instabilities in the Fermi surface spin liquid states discussed above. Here we want to address this issue. We now know that besides the trimer state, there are actually two spin liquid states--zero-flux spin liquid state and $\pi$-flux spin liquid state. Both of these spin liquid states are gapless and contain a single or multiple parton Fermi surfaces. Focusing on these regimes, we take the pairing mechanism into consideration and the trial Hamiltonian becomes
\begin{eqnarray}\label{MF_H:pairing}
\nonumber H_{trial}&=&-\sum_{\la j k \ra} \sum_{\alpha, \beta} \bigg{[} \bigg{(}  t_{jk}^\phi \delta^{\alpha \beta} f^{\alpha \dagger}_j f^{\beta}_k +  \Delta^{\alpha\beta}_{jk} f^{\alpha\dagger}_j f^{\beta \dagger}_k \bigg{)} +\Hc \bigg{]}\\
&& -\sum_{j}\sum_{\alpha}\mu_j f^{\alpha\dagger}_jf^{\alpha}_k,~
\end{eqnarray}
with the constraint, $n^{\alpha}_j = \la f^{\alpha\dagger}_j f^{\alpha}_j\ra = 1/3$ and $t^\phi_{jk} = \pm t_{jk}$ for flux $\phi=0/\pi$. For clarity, from now on we will replace the bond labeling $\la j k \ra$ by $({\bf r}, {\bf r} + \vec{e}_{\nu})$, with ${\bf r}$ running over all lattice sites and $\vec{e}_{\nu = 1,~2,~3}$ are shown in Fig.~\ref{triangular_lattice}. We abbreviated the sum over all $\vec{e}_{\nu = 1,~2,~3}$ as $\vec{e}$. 

In these regimes, the optimal state are uniform flux states which suggests uniform trial hopping amplitude $t_{jk} \simeq t \equiv 1$ and the uniform expectation values of hopping functions, $\la f^{\alpha\dagger}({\bf r})f^{\beta}({\bf r}+{\vec{e}}_\nu )\ra =\delta^{\alpha \beta} \chi^{\alpha *} (\vec{e}_\nu).$ Furthermore, in these regimes, the hopping functions, $\chi^\alpha(\vec{e}_\nu)$, should be real and are numerically confirmed. Below, we consider two pairing cases: Case (1) corresponds to pairing within the same flavor of fermions. This requires the orbital angular momentum quantum number to be $l=1,~3,~...$ corresponding to $p_x + i p_y$,  $f$-wave, ...pairing states; Case (2) corresponds to BCS-type pairing with different flavor of fermions. This pairing requires $l = 0,~2,~...$ corresponding to $s$-wave, $d_x + i d_y$,... pairing states.\cite{Serbyn2011, Bieri2012} Below, we discuss the mathematical set up for each case separately.\\

{\it Case (1): pairing ansatz with $\Delta^{\alpha \beta} = \delta^{\alpha \beta} \Delta^{\alpha}$}.\\

The pairing functions we consider are $\la f^{\alpha}({\bf r})f^{\beta}({\bf r}+\vec{e}_\nu)\ra = \delta^{\alpha \beta} \Delta^{\alpha}(\vec{e}_\nu).$  By symmetry arguments, the model conserves spatial rotational symmetry and as already explained above, the angular momentum of pairing functions should be odd, $\mathit{l} = 1,~3,~....$ corresponding to $p_x + i p_y$, $f$-wave pairing,..... The ansatz can be further simplified to be $\chi^{\alpha}(\vec{e}_1) = \chi^{\alpha}(\vec{e}_2)=\chi^{\alpha}(\vec{e}_3)\equiv \chi^{\alpha},~\label{rotinv_chi}$ and $\Delta^{\alpha}(\vec{e}_3)=\Delta^{\alpha}(\vec{e}_2)e^{i \mathit{l} \cdot \theta} =\Delta^{\alpha}(\vec{e}_1)e^{i \mathit{l} \cdot 2 \theta}\equiv \Delta^{\alpha} e^{i \mathit{l}\cdot 2 \theta}$, where $\theta$ is $2 \pi/3$ shown in Fig.~\ref{triangular_lattice}. In addition, because the vectors $\vec{f} = \{ f^{x}, f^{y}, f^{z} \}$ and $\vec{f^{\dagger}} = \{ f^{x\dagger}, f^{y \dagger}, f^{z \dagger} \}$ transform as a three-dimensional vectors under spin rotation, we expect $\chi^x = \chi^y=\chi^z\equiv \chi$ and $\Delta^{x}=\Delta^{y}=\Delta^{z}\equiv \Delta$. In this pairing scheme, the $SU(3)$ is broken down to $SO(3)$. It is straightforward to diagonalize the mean-field Hamiltonian
\begin{eqnarray}
H_{trial} = \sum_{\alpha}\sum_{{\bf k}\in {\bf B.Z.}} E_{\alpha}({\bf k})a^{\alpha \dagger}({\bf k})a^{\alpha}({\bf k}),~\end{eqnarray}
where $a^\alpha ({\bf k})$ are Bogoliubov quasiparticles satisfying the transformation
\begin{eqnarray}
\begin{pmatrix}
f^{\alpha}({\bf k})\\
f^{\alpha\dagger}(-{\bf k})
\end{pmatrix}
=
\begin{pmatrix}
u^{\alpha}_{{\bf k}} & - v^\alpha_{{\bf k}}\\
v^{\alpha *}_{{\bf k}} & u^{\alpha *}_{{\bf k}}
\end{pmatrix}
\begin{pmatrix}
a^{\alpha}({\bf k})\\
a^{\alpha\dagger}({\bf k})
\end{pmatrix},~
\end{eqnarray}
with 
\begin{eqnarray}
\nonumber && |u^{\alpha}_{\bf k}|^2 = \frac{1}{2} \left[ 1+ \frac{\xi_{\alpha}({\bf k})}{E_{\alpha}({\bf k})} \right], \\
&& |v^{\alpha}_{\bf k}|^2 = \frac{1}{2} \left[ 1 - \frac{\xi_{\alpha}({\bf k})}{E_{\alpha}({\bf k})}\right].\label{BCS_uv}
\end{eqnarray}
The ground state can be written as 
\begin{eqnarray}\label{MF_GS}
| GS \ra = \prod_{\alpha=x,y,z} \prod_{{\bf k}} \left[ u^{\alpha}_{{\bf k}} + v^{\alpha}_{{\bf k}}f^{\alpha \dagger}({\bf k})f^{\alpha\dagger}(-{\bf k})\right]|vac \ra,~
\end{eqnarray}
where we define
\begin{eqnarray}
&& \xi_{\alpha}({\bf k})\equiv - \sum_{{\vec{e}}} 2 \cos({\bf k}\cdot {\vec{e}}) - \mu,~\label{BCS_xi}\\
&& \tilde{\Delta}_{\alpha}({\bf k})\equiv  \sum_{{\vec{e}}}  i \Delta \sin({\bf k}\cdot {\vec{e}}),~\\
&& E_{\alpha}({\bf k})=\sqrt{(\xi_{\alpha}({\bf k})/2)^2+|\tilde{\Delta}_{\alpha}({\bf k})|^2}.~
\end{eqnarray}
We can see there are three degenerate bands in this case.\\

{\it Case (2):pairing ansatz with} $\Delta^{\alpha \alpha}=0,~\Delta^{\alpha\beta}|_{\alpha \not= \beta} \not= 0$.\\

We consider the pairing function of the form, $\la f^{\alpha}({\bf r})f^{\beta}({\bf r} + \vec{e}_{\nu})\ra_{\alpha \not= \beta} =\Delta^{\alpha \beta}(\vec{e}_{\nu})$. It seems there are three pairing functions we need to consider, $\Delta^{xy}$, $\Delta^{yz}$, and $\Delta^{zx}$, but in this SU(3) symmetric model, we can perform a global gauge transformation to make $\Delta^{yz},~\Delta^{zx} = 0$ as long as the length of the vector formed by $\Delta$s is conserved, $|\Delta^{xy}|^2 + |\Delta^{yz}|^2 + |\Delta^{zx}|^2 = \Delta^2 =$ constant. \cite{Honerkamp_BCS, Honerkamp2004, He2006, Cherng2007, Chung2010, OHara2011} If this pairing state is energetically favored, there is always one flavor of gapless fermions in this SU(3) system which we choose to be $f^z$.

From now on, we will set $\Delta^{yz} = \Delta^{zx} = 0$ and $\Delta^{xy} = \sqrt{3} \Delta$. In this gauge choice, the symmetry breaking process is more apparent. The symmetry breaking is $SU(3) \rightarrow SU(2) \otimes U(1)$. The $SU(2)$ symmetry is generated by the psudo-spin doublet $f^{x}$ and $f^{y}$ and the $U(1)$ is generated by the gapless $f^z$ fermion. After Bogoliubov transformation, the ground state is 
\begin{eqnarray}
|GS \ra = && \prod_{{\bf k}} [ u^{xy}_{{\bf k}} + v^{xy}_{{\bf k}} f^{x \dagger}({\bf k}) f^{y \dagger}(-{\bf k}) ] |vac\ra \otimes \\
&& \otimes \sum_{\xi^z_{{\bf k}} <0} f^{z \dagger}({\bf k}) | vac \ra, \label{freefz}
\end{eqnarray}
where the second part, Eq.~(\ref{freefz}), is the wave function of the free $f^z$ fermion, and $u^{xy}_k$ and $v^{xy}_k$ have the same expression as in Eq.~(\ref{BCS_uv}) with $\xi^{x}({\bf k})=\xi^{y}({\bf k}) = \xi^{z}({\bf k})\equiv \xi({\bf k})$ the same to Eq.~(\ref{BCS_xi}) and 
\begin{eqnarray}
&& \tilde{\Delta}^{xy}({\bf k}) \equiv \sum_{\vec{e}} \sqrt{3} \Delta \cos({\bf k} \cdot \vec{e}),\\
&& E^{xy}({\bf k}) = \sqrt{(\xi({\bf k}))^2 + | \tilde{\Delta}^{xy}({\bf k}) |^2}.
\end{eqnarray}
We can see in the SU(3)-symmetric point, the energy bands always show one gapless branch corresponding to one flavor of gapless fermions. 

With the two pairing ansatzes above, we focus on the regimes of the zero-flux spin liquid state and the $\pi$-flux spin liquid state in Fig.~\ref{MFSL_pdg}. We again perform full Wick contractions and ignore the pure constant density terms. The trial energy expression after Wick contractions is too complex to write out explicitly. We test all different pairing ansatzes above and numerically calculate the trial energies with the optimal pairing functions $\Delta$ in the uniform-flux spin liquids regime on the triangular lattice with $300 \times 300$ sites. 

\begin{figure}[t]
\includegraphics[width=\columnwidth]{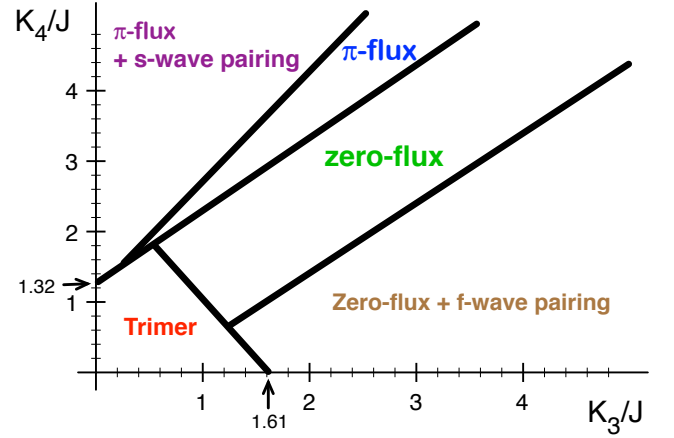}
\caption{The mean-field phase diagram with pairing instability. The lines are the tentative boundaries between different states. We note that the line separating the $\pi$-flux and zero-flux spin liquid states and the line between zero-flux and trimer state are the same as shown in Fig.~\ref{MFSL_pdg}. Compared with Fig.~\ref{MFSL_pdg}, the phase diagram shows that the zero-flux spin liquid state has a possible pairing instability toward a $f$-wave gapless (nodal) spin liquid state, and the $\pi$-flux spin liquid state has a possible pairing instability toward a $s$-wave spin liquid state with two flavors of fermions paired up while one flavor of fermions remain gapless.
}
\label{pairing_pdg}
\end{figure}

The result is summarized in Fig.~\ref{pairing_pdg}. We find that roughly in the regime where the FM three-site ring exchanges slightly dominant, the zero-flux spin liquid state (gapless spin liquid with parton Fermi surface) has the pairing instability toward a $f$-wave gapless (nodal) spin liquid. When the  four-site ring exchange $K_4$ is strong enough, the pairing instability can be suppressed and the optimal state is the zero-flux spin liquid with Fermi surfaces. However, when we keep increasing $K_4$, the optimal state becomes the uniform $\pi$-flux spin liquid state. Interestingly, in the $\pi$-flux  spin liquid state, roughly when the four-site ring exchanges dominant, such a $\pi$-flux spin liquid state has the pairing instability toward an exotic $s$-wave spin liquid state with $f^x$ pair with $f^y$, which forms a psudo-spin singlet, while $f^z$ remains gapless.  We note that we numerically find no pairing instability of the uniform spin liquid states toward $p_x + i p_y$ spin liquid states or $d_x + i d_y$ spin liquid states in the focusing regime where $K_3, K_4 >0$.
\section{Discussion}\label{Sec:Discussion}
We study the SU(3) ring-exchange model with ``FM`` three site ring exchanges and  ``AFM`` four-site ring exchanges. We first use the site-factorized ansatz to study the model and find the three-sublattice ordered states, FM states in a large regime of $K_3-K_4$ parameter regime. 

In the slave-fermion trial states studies, we find the main competing states are the trimer state, uniform $\pi$-flux spin liquid state and the zero-flux spin liquid. The trimer state is strongly suppressed by increasing the strength of the ring exchanges. We also find that the zero-flux state has a possible pairing instability toward a $f$-wave gapless (nodal) spin liquid and the $\pi$-flux spin liquid state has a pairing instability toward an exotic $s$-wave spin liquid state with $f^x$ pairing with $f^y$ fermions while $f^z$ fermions remain gapless. 

We note that it is not legitimate to compare the phase diagram obtained from the site-factorized state studies, Fig.~\ref{site_factorized_pdg}, with those obtained from the mean-field slave-fermion trial state studies, Figs.~\ref{MFSL_pdg}-\ref{pairing_pdg}. It is more appropriate to compare the energetics of the site-factorized states with those of the Gutzwiller-projected states, which can be obtained by performing Variational Monte Carlo (VMC) studies and Gutzwiller projection on the mean-field states we obtain in this paper, \cite{WenPSG} which is beyond the scope of the paper. Besides, in the slave-fermion trial states studies, we only focus on the non-magnetic trial states and fix the number density of each flavor to be equal, $n^{\alpha}_j = \la f^{\alpha\dagger}_j f^{\alpha}_j \ra =1/3$.  In principle, we can also consider the magnetic ordered states by making the number density of each flavor per site different as, for example, $\la f^{z\dagger}_j f^{z}_j\ra = 1$, and $\la f^{x\dagger}_j f^{x}_j\ra=0=\la f^{y\dagger}_j f^{y}_j\ra$. However, similar to the previous discussion, we still need to perform Gutzwiller projection on all these fermionic mean-field states (magnetic ordered states and non-magnetic states) in order to compare the energies of different states even within the slave-fermion trial studies.

For further clarifying in which regime the spin liquid states are more robust and qualitatively relate the phase diagrams of the site-factorized studies and mean-field slave-fermion trial state studies, Fig.~\ref{site_factorized_pdg} and Figs.~\ref{MFSL_pdg}-\ref{pairing_pdg}, we follow Ref.~\onlinecite{Bauer2012} to perform linear flavor wave theory (LFWT), an extension of spin wave theory to SU(N) model, on the three-sublattice ordered state. \cite{Tsunetsugu2006, note:LFWT} We find that the energy per site after taking the quantum fluctuations into account is 
\begin{eqnarray}
E_{3-sub}|_{LFWT} \simeq -0.6295 J + 0.5462 K_3 + 0.2568 K_4,~~~
\end{eqnarray}
which is consistent with the result in Ref.~\onlinecite{Bauer2012} at $K_3,~K_4\rightarrow 0$. We can see that even though the quantum fluctuations lower the (two-site exchange) Heisenberg energy from $0$ to $-0.63 J$, the ring exchange energies increase. At strong $K_3$ and $K_4$, it is expected that quantum fluctuations eventually destroy the three-sublattice ordered state. On the other hand, since FM state in Fig.~\ref{site_factorized_pdg} is an exact eigenstate of the SU(3)-ring exchange Hamiltonian, the FM is a much stable phase and the interesting quantum spin liquid states are unlikely to arise in the FM regime. Therefore, in this SU(3)-ring exchange model, we expect that the interesting quantum spin liquid states can arise in the large $K_3$, $K_4$ parameter regime of the three-sublattice ordered state. 

The gapless spin liquid states have different properties and can be distinguished, at least in the mean-field picture. For clarity, below we will consider specifically the spin correlations, $\la \vec{S}_j \cdot \vec{S}_k \ra_{conn}$ and the (nematic) correlations of diagonal elements of tranceless quandrupolar tensor, $\la \mathcal{Q}^{\alpha \alpha}_j \mathcal{Q}^{\alpha}_k \ra_{conn} $ with $\mathcal{Q}^{\alpha \beta}_{j} =  (S^\alpha_j S^\beta_j + S^{\beta}_j S^{\alpha}_j )/2 - 2\delta^{\alpha \beta}/3$. Above we defined the connected correlations $\la \mathcal{O}^{\dagger}_j \mathcal{O}_k\ra_{conn} \equiv \la \mathcal{O}_j \mathcal{O}_k\ra - \la \mathcal{O}^{\dagger}_j \ra \la \mathcal{O}_k \ra$, with $\mathcal{O}_j = \vec{S}_j$ or $\mathcal{Q}_j$. Before jumping into the discussions of the properties of different spin liquid states, we first note that the correlation functions of the diagonal elements of the quadrupolar tensor can be expressed as
\begin{eqnarray}
\la \mathcal{Q}^{\alpha \alpha}_j \mathcal{Q}^{\alpha \alpha}_k \ra_{conn} = \frac{4}{9}- \left|\la f^{\alpha \dagger}_j f^{\alpha}_k \ra \right|^2,~
\end{eqnarray}
where above we used the identity, $\mathcal{Q}^{\alpha \beta}_j = \delta^{\alpha \beta} - f^{\alpha \dagger}_j f^{\alpha}_j$ with the constraint $\sum_\alpha f^{\alpha\dagger}_j f^{\alpha}_j =1$.\cite{Bieri2012} We can see the first constant is universal for different spin liquid states, but the second contribution is qualitatively different in different spin liquid states can be used for characterizing different spin liquid state. For simplicity of discussion, we define $\la \mathcal{Q}^{\alpha \alpha}_j \mathcal{Q}^{\alpha \alpha}_k \ra_{-}$. We will also discuss the thermal properties of each spin liquid state.\\

{\it Properties of uniform-flux (U(1)) spin liquid states:} The uniform zero-flux and $\pi$-flux spin liquid states are both U(1) (Fermi-surface) spin liquid states. The zero-flux spin liquid possesses a single parton Fermi pocket in the center and the $\pi$-flux spin liquid possesses two parton Fermi pockets near the hexagonal Brillouin zone. In these two different U(1) spin liquid states, 
\begin{eqnarray}
\nonumber \la \vec{S}_j \cdot \vec{S}_k \ra^{mf}_{conn} &\sim& \la \mathcal{Q}^{\alpha \alpha}_j \mathcal{Q}^{\alpha \alpha}_k \ra^{mf}_{-} \sim\\ 
& \sim&  \frac{1+ \cos{[(\bf{k}_{FR} - \bf{k}_{FL})\cdot(\bf{r}_k - \bf{r}_j)]}}{|\bf{r}_k - \bf{r}_j|^3},
\end{eqnarray}
where the $\bf{k}_{FR/L}$ represent the momenta of the right patch and the left patch of the Fermi surface for an observable direction. Since the two Fermi surface spin liquid states have different geometric information of Fermi surfaces, the corresponding wave vectors are quantitatively different and can be detected by studying the corresponding spin structure factors. Since the uniform-flux spin liquid states have gapless parton Fermi surface(s), we expect to see linear-temperature dependent specific heat $(C_v \propto T)$ and thermal conductivity $(\kappa \propto T)$.\\

{\it Properties of $f$-wave gapless (nodal) spin liquid state}: This spin liquid state possesses gapless nodal points. Interestingly, even though the pairing break the original SU(3) symmetry, the SO(3) rotational symmetry related to the spin rotation is still preserved (because of the fact that $\vec{f}$ and $\vec{f}^\dagger$ with $\vec{f} \equiv (f^x, f^y, f^z)$ transform as three-dimensional vectors). As far as the low-energy physics is concerned, this spin liquid state possesses 
\begin{eqnarray}
\la \vec{S}_j \cdot \vec{S}_k \ra^{mf}_{conn} &\sim& \la \mathcal{Q}^{\alpha \alpha}_j \mathcal{Q}^{\alpha \alpha}_k \ra^{mf}_{-}  \sim \frac{1 + \cos[\bf{k}_c \cdot (\bf {r}_k - {\bf r}_j)]}{|\bf{r}_k - \bf{r}_j|^4},~~~~~~
\end{eqnarray}
with ${\bf k}_c$ being the vectors connecting different Dirac points and we ignore all the pre-factors of each term in the above equation. Since this spin liquid state contains gapless nodal points, it should possess square-temperature specific heat ($C_v \propto T^2$).

{\it Properties of the exotic $s$-wave spin liquid state}: This spin liquid state possesses {\it short-ranged } spin correlations due to the superconducting gap in the $x$-$y$ channel. In order to detect the gaplessness of such a spin liquid state, we can measure the correlation functions of different diagonal elements of the quadrupolar tensor. For example, the $\mathcal{Q}^{zz}$ correlation will show the power-law behavior
\begin{eqnarray}
\la \mathcal{Q}^{zz}_j \mathcal{Q}^{zz}_k \ra^{mf}_- \sim \frac{1 + \cos[ ( \bf{k}^z_{FR} - \bf{k}^z_{FL})\cdot (\bf{r}_k - \bf{r}_j )]}{|\bf{r}_k - \bf{r}_j |^3},
\end{eqnarray}
 with $\bf{k}^z_{FR/L}$ being the momenta of the right patch and the left patch of the $f^z$ Fermi surface. But the correlations related to $\mathcal{Q}^{xx}$ and $\mathcal{Q}^{yy}$ exponentially decay. As for the thermal properties, we expect to see $C_v\propto T$ and $\kappa\propto T$ due to the gapless $f^z$ parton Fermi surface.

From the perspective of numerics, since there have been DMRG and iPEPS studies on the SU(3) Heisenberg model of three-flavor fermions on the triangular lattice which found the three-sublattice ordered state. \cite{Bauer2012} We suggest possibly the interesting zero-flux spin liquid state be also detected in the DMRG and iPEPS studies on the SU(3) ring-exchange model. 

From the view of cold atom experiments. Recently, the cold atom experiment demonstrated a method to be able to add an artificial tunable gauge potential to the system. \cite{Struck2012} With the tunable gauge potential, it may be possible to tune the sign of the three-site ring exchanges from FM to AFM. In that case, for strong four-site ring exchange, the main competing states are still the zero-flux and $\pi$-flux spin liquid states, with the trial energies similar to Eqs~(\ref{MF_zeroflux:num_val})-(\ref{MF_piflux:num_val}) with $K_3 <0$. However, for small $K_4$, the main competing slave-fermion trial states are the uniform $\pi/3$-flux spin liquid state with the trial energy, $E^{MF}_{\phi=\pi/3}\simeq -0.8992 J + 0.8343 K_3 +0.02443 K_4$, and the $\pi$-flux spin liquid state. It may be interesting to explore the regime with AFM three-site ring exchange with $K_4 \sim 0$ and it may be possible to see a phase transition between these two spin liquid states by manipulating the artificial gauge potential in experiment.

Another interesting theoretical outlook is how the phase diagram, Fig.~\ref{pairing_pdg}, evolves if we perturb the model away from the SU(3)-symmetric point. In this way, the model can be connected to other models \cite{Serbyn2011, Xu2012, Bieri2012} which can explain the gapless spin-1 spin liquids possibly realized in Ba$_3$NiSb$_2$O$_9$ \cite{Cheng_spin1SL} or other theoretical spin-1 models. \cite{Papanicolaou1988, Bhattacharjee2006, Tsunetsugu2006, Ng2010, Grover2011} 

In the present model studies, we treat the parameters $J$, $K_3$, and $K_4$ as three independent parameters. However, from the perspective of the perturbation studies on the Hubbard-to-Mott insulator transition, as pointed out in the beginning, the parameters $J$, $K_3$, and $K_4$ are actually not independent to each other. According to the mean-field phase diagram, Fig.~\ref{MFSL_pdg}, the interesting gapless spin liquid states are more likely to arise in the parameter regime with $K_3/J$, $K_4/J$ of the order one or greater, which means the system on the insulating side is closer to the transition point. In this regime, the perturbation theory breaks and higher order terms need to be included, which not only makes the theoretical analysis out of control but also points out the possibility that the experimental realization of the interesting gapless spin liquid states is out of reach. In order to have a well-controlled theoretical analysis to get access to such interesting gapless spin liquid states and to possibly shad light on the experimental realization in the cold atom systems, we would like to study a Fermi-Hubbard-type model on a two-leg ladder system with on-site or more extended repulsion in the future, similar to the analysis outlined in Ref.~\onlinecite{Lai10}. 
\acknowledgments
The author would like to thank deeply Olexei I. Motrunich for his suggestions and critically read the initial manuscript. The author thanks Kun Yang, Yuan-Ming Lu, Gang Chen, Maksym Serbyn, Samuel Bieri, and Frederic Mila for helpful discussion. The author is supported by National Science Foundation under Grant No. DMR-1004545 and DMR-0907145. The author also would like to thank KITP where the research is completed and is supported in part by the National Science Foundation under Grant No. NSF PHY11-25915.

\bibliography{biblio4SU(3)}

\begin{thebibliography}{46}
\bibitem [{Mar()}]{Mark:helical_state}%
  \BibitemOpen
  \href@noop {} {}\bibinfo {note} {The four-sublattice ordered state can be
  possibly analyzed numerically by considering the helical states. The helical
  state can be constructed by first choosing the site-factorized vectors
  $\vec{\mathcal{X}}_l$ and $\vec{\mathcal{X}}_{l+1}$ of two consecutive
  stripes (in the basis of $\{ |x \ra, |y\ra, |z\ra\}$, and we can write the
  next one as $\vec{\mathcal{X}}_{l+2} = \cos\theta \vec{\mathcal{X}}_l +
  \sin\theta
  \vec{\mathcal{X}}_{l}\times\vec{\mathcal{X}}^{\dagger}_{l+1}$.}\BibitemShut
  {Stop}%
\end{thebibliography}
\end{document}